\documentclass[final]{aipproc}
\layoutstyle{6x9}

\usepackage{graphicx}% Include figure files
\usepackage[usenames]{color}

\newcommand{\onepion}{one pion~}

\newcommand{\cm}{\; \mathrm{cm}}
\newcommand{\GeV}{\; \mathrm{GeV}}

\begin{document}

\title{Neutrino nucleus reactions at high energies within the GiBUU model}
\author{O. Lalakulich}{
  address={Institut f\"ur Theoretische Physik, Universit\"at Giessen, Germany}
}
\author{K. Gallmeister}{}
\author{T. Leitner}{}
\author{U. Mosel}{}

\begin{abstract}
The GiBUU model, which implements all reaction channels relevant at medium neutrino energy,
is used to investigate the neutrino and 
antineutrino scattering on iron. Results for integrated cross sections are compared with NOMAD and MINOS data.
It is shown, that final state interaction can noticeably change the spectra of the outgoing hadrons.
Predictions for the Miner$\nu$a experiment are made for pion spectra, averaged over 
NuMI neutrino and antineutrino fluxes.
\end{abstract}

\date{\today}

\maketitle

\section{Introduction}

Neutrino and antineutrino scattering on nuclei for neutrino energies above $30 \GeV$ was studied 
in several experiments starting from the 80s (see \cite{Conrad:1997ne,Tzanov:2009zz} for a review). 
Based on muon detection, integrated cross sections were measured with combined precision of 2\%;  
they grow linearly with energy,  which agrees with the predictions of the quark parton model.
Double differential cross sections with respect to muon variables were also
measured, allowing one to extract the nucleon structure functions. 

At lower energies the neutrino reactions are not so easy to model, because of
the overlapping contributions from QE scattering, resonance production and 
background processes.
Another difficulty is in choosing an appropriate way to describe the onset  of the DIS processes, 
the so-called Shallow Inelastic Scattering. 
This requires complex approaches that take all of the relevant channels into account.
The lack of data at intermediate energies has long been an obstacle for a serious test of such approaches.

Up-to-date and coming experimental results are changing the situation.
Recently NOMAD reported the inclusive neutrino cross section for $E_\nu>4.6 \GeV$ with an accuracy
of at least $4\%$ \cite{:2007rv}. MINOS reported both neutrino (for $E_\nu>3.48\GeV$) 
and antineutrino (for $E_{\bar\nu}>6.07\GeV$) cross sections with a comparable precision \cite{Adamson:2009ju}.
The Miner$\nu$a experiment intends to perform measurements on
Plastic (CH), Iron, Lead, Carbon, Water and liquid Helium targets in the NuMI beam, 
which would directly  allow to compare nuclear effects on various nuclei.  
Besides muon detection, this experiment will also be able to resolve various final states
by identifying the tracks of the outgoing hadrons. 

Here we use the GiBUU transport model \cite{gibuu,Buss:2011mx} to study neutrino and antineutrino scattering
on iron. Our results  are compared
with the recent MINOS and NOMAD data.  Predictions are also made for the spectra of the outgoing particles, 
which can be measured by the Miner$\nu$a experiment.
The calculations are done without any fine tuning to the 
data covered here with the default parameters as used in the GiBUU framework.

\section{GiBUU transport model. Various neutrino processes and problem of double counting}

The GiBUU model was initially developed 
as a transport model for nucleon-, \mbox{nucleus-,}  pion-, and electron- induced reactions from
some hundreds MeV up to tens of GeV. Several years ago neutrino-induced interactions were
also implemented for the energies up to a few GeV. 
Recently the code was extended to describe also the DIS processes in neutrino reactions.
Thus, we can study all kind of elementary collisions on all kind of nuclei within a unified framework. 
The model is based on well-founded theoretical ingredients and has been tested against various nuclear reactions.
For a detailed review of the GiBUU model see \cite{Buss:2011mx}.

GiBUU describes all processes relevant at medium energies,
Our approach to quasielastic (QE) scattering, resonance (RES) production and background (BG) processes 
is described in \cite{Leitner:2006ww,Leitner:2008ue}. 
The DIS scattering is  included   as \textsc{Pythia} simulation.

In the region of the shallow inelastic scattering, that is at moderate 
invariant masses, $1.6\GeV<W<2.0\GeV$,  there is a potential problem of double counting.  
Here the same physical  events can be considered as originating from decays of 
high mass baryonic resonances  or from DIS.
In the GiBUU code we use the ansatz, that both RES, BG and DIS processes contribute in this region. 
The RES and BG contributions are smoothly switched off
above $W_{\rm RES-2}$ and DIS contribution is switched on above $W_{\rm{DIS-1}}$:
\begin{equation}
 \sigma_{\rm RES}=\sigma_{\rm RES}(W)\frac{W_{\rm RES-2}-W}{W_{\rm RES-2}-W_{\rm RES-1}}
\qquad
 \sigma_{\rm DIS}=\sigma_{\rm DIS}(W)\frac{W-W_{\rm DIS-1}}{W_{\rm DIS-2}-W_{\rm DIS-1}}
\label{Wcuts}
\end{equation}
The default parameters are: $W_{\rm RES-1}=2.0\GeV$, $W_{\rm RES-2}=2.05\GeV$, $W_{\rm DIS-1}=1.6\GeV$, and $W_{\rm DIS-2}=1.65\GeV$.
With this choice, the DIS events become noticeable at neutrino energies around $3\GeV$.  
This choice is mainly motivated by the comparison with the electroproduction data  \cite{Christy:2007ve}.

In essence, the DIS processes below $W_{\rm RES-2}$ account for the
resonances whose electromagnetic properties are not known  and for the 
non-resonant processes giving a few mesons in the final state beyond the \onepion{} background.

\section{Integrated cross sections.}

As already mentioned, the DIS cross section grows linearly with energy. 
So at high neutrino energies the data are conveniently presented as cross section  per energy $\sigma_{tot}/E_\nu$.  
The world average values are given \cite{Amsler:2008zzb} for isoscalar target:
$\sigma_{tot}/E_\nu (\nu) = 0.667\pm 0.014 \cdot 10^{-38} \cm^2/\GeV$ for neutrinos
and $\sigma_{tot}/E_\nu (\bar\nu) = 0.334\pm 0.008 \cdot 10^{-38} \cm^2/\GeV$ for antineutrinos.

\begin{figure}[hbt]
\begin{minipage}[c]{0.48\textwidth}
\includegraphics[angle=-90,width=\textwidth]{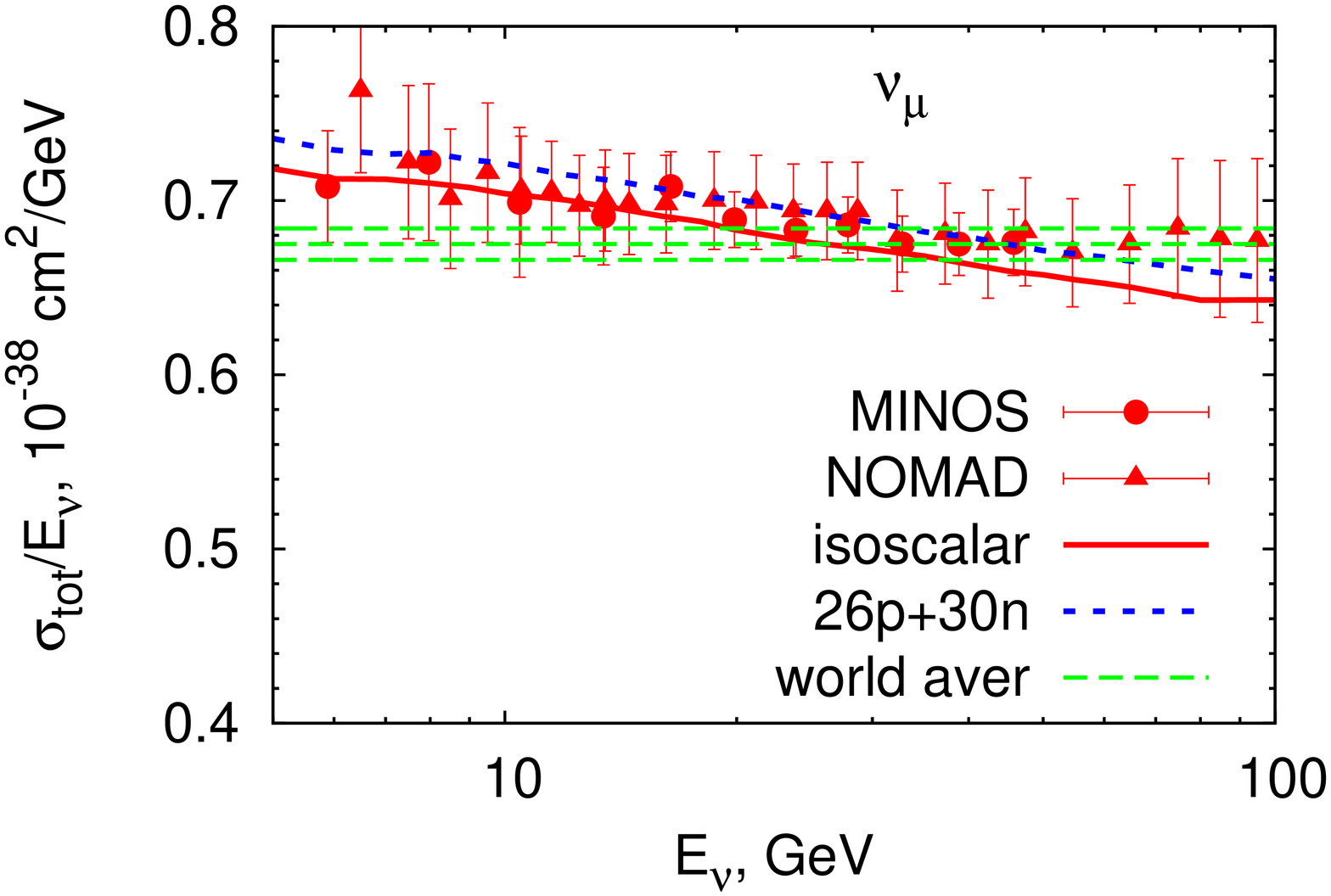}
\end{minipage}
\hfill
\begin{minipage}[c]{0.48\textwidth}
\includegraphics[angle=-90,width=\textwidth]{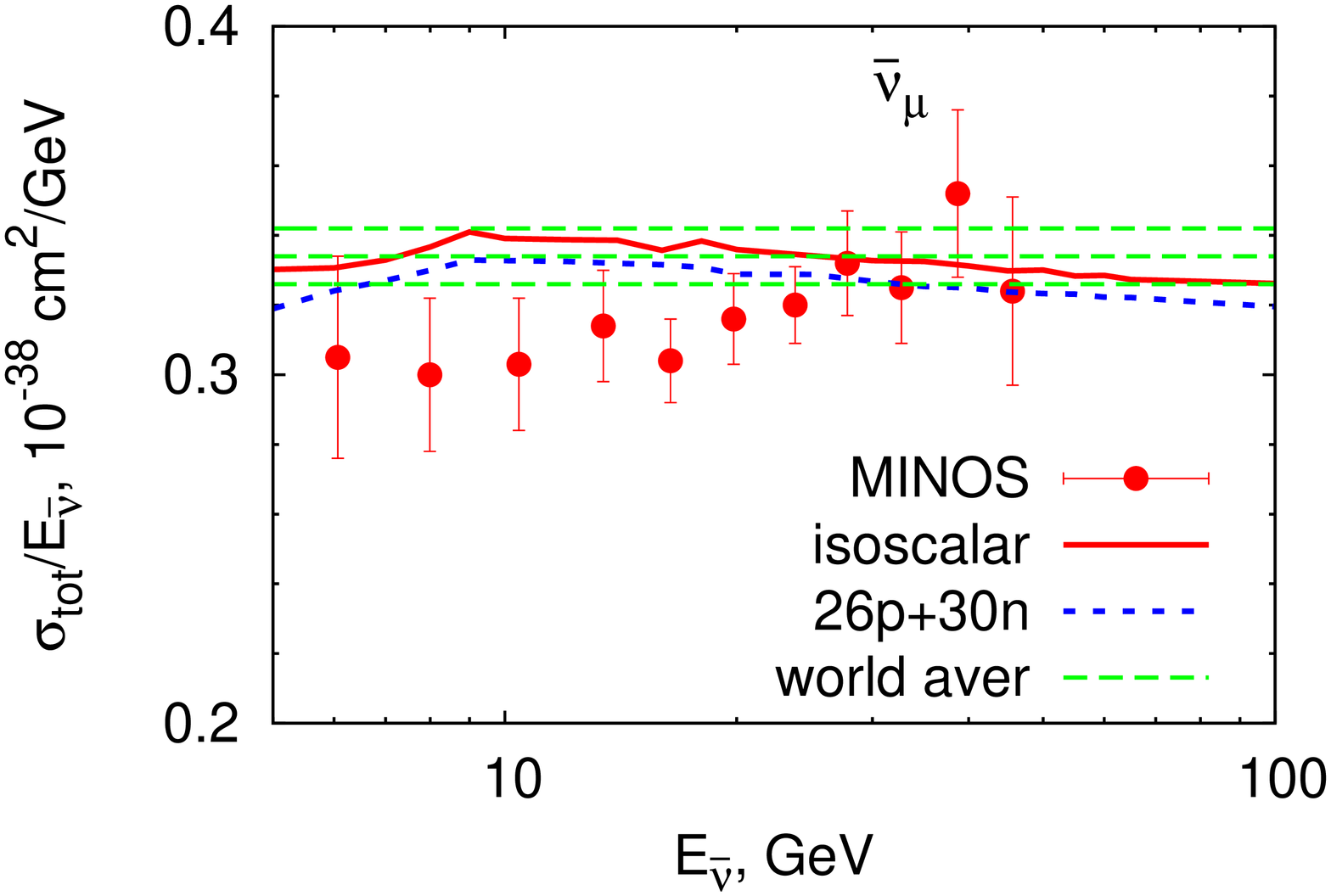}
\end{minipage}
\label{fig:compare-isoscalar-free}
\caption{(Color online) Total elementary cross section per nucleon for neutrino (left) and antineutrino (right) 
induced reactions. 
The world average values \protect\cite{Amsler:2008zzb} and their errorbands are also shown as long-dashed green lines.
Experimental data are from \protect\cite{:2007rv,Adamson:2009ju}.}
\end{figure}

Fig.~\ref{fig:compare-isoscalar-free} shows the isoscalar cross section together with the 
free cross section for the iron composition, that is for the 26 protons and 30 neutrons 
(46.4\% protons, 53.6\% neutrons). Recent MINOS data for iron target as well as NOMAD data for 
an isoscalar target are also shown. 

This figure shows that the decreasing slope of our curves for the neutrino cross section
is in agreement with that of the data. This slope is not taken into account in deriving the world-average
value, where it was assumed to be negligible.  Notice, that such a comparison with the data is meaningful,
only if nuclear corrections are neglected.

\begin{figure}[hbt]
\begin{minipage}[c]{0.48\textwidth}
\includegraphics[width=\textwidth]{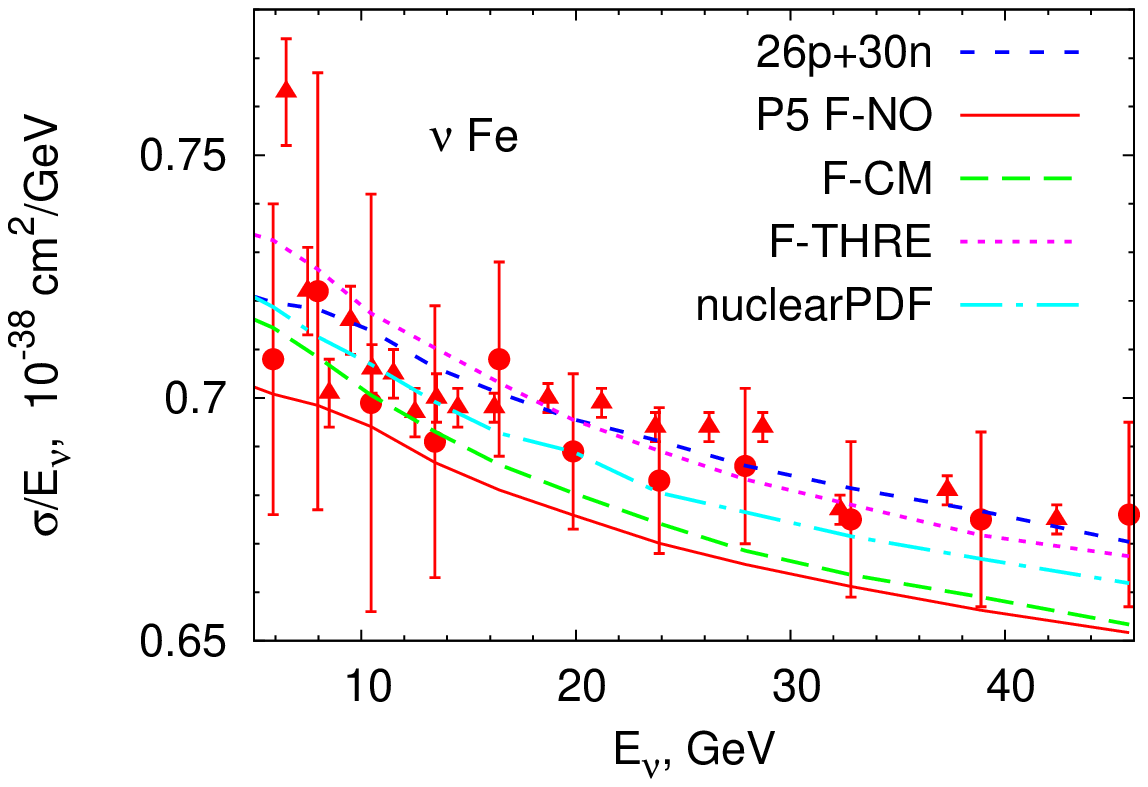}
\end{minipage}
\hfill
\begin{minipage}[c]{0.48\textwidth}
\includegraphics[width=\textwidth]{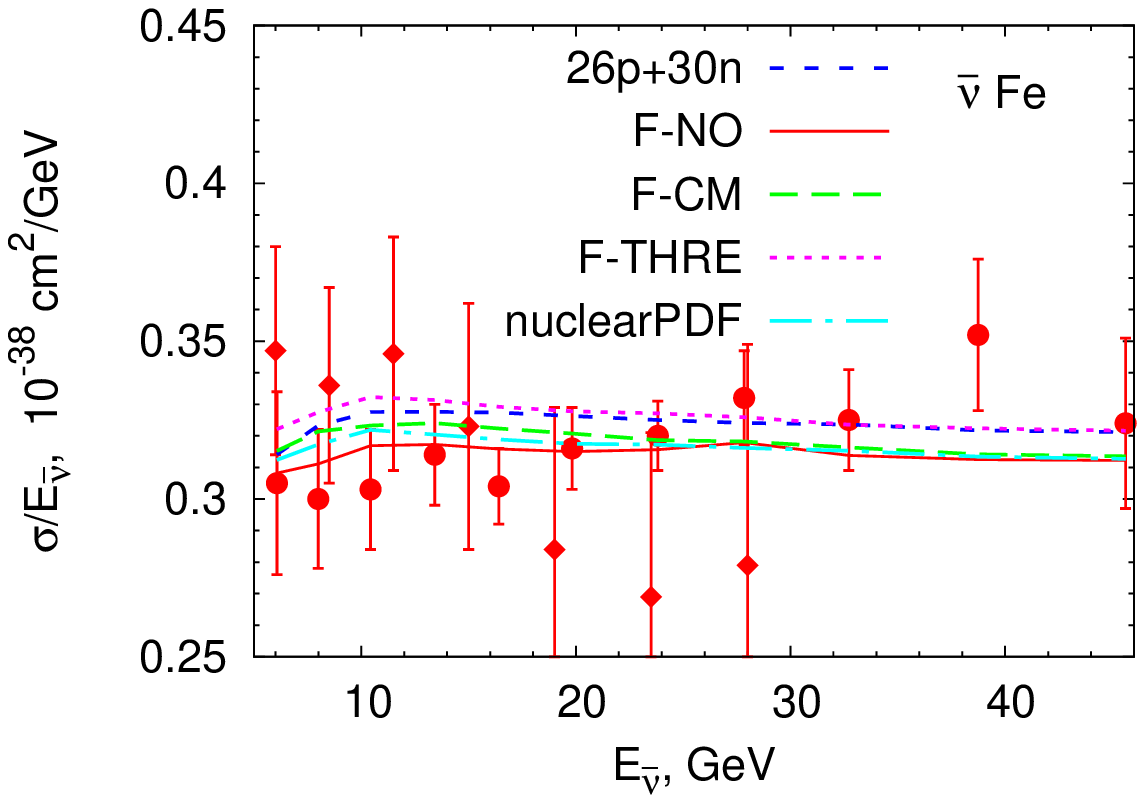}
\end{minipage}
\caption{(Color online) Total cross section per nucleon for neutrino (left) and antineutrino (right) 
induced reactions on iron target. Various prescriptions to include nuclear effects in DIS are compared.
MINOS data (full circles) are from \protect\cite{:2007rv,Adamson:2009ju}, 
IHEP-JINR data (full diamonds) are from \protect\cite{Anikeev:1995dj}.}
\label{fig:compare-nuclear}
\end{figure}

The actual value of such corrections for neutrino reactions is not known so far, 
because of both experimental inaccuracies and difficulties in the theoretical description. 
On one hand, nuclear parton distributions, based on electromagnetic scattering data and intended for description
of both charged lepton and neutrino reactions, were introduced. For a review and a list 
of recent parametrization see, for example, \cite{Hirai:2009mq}.
On the other hand,
recent investigation \cite{Schienbein:2007fs,Kovarik:2010uv} showed, that in neutrino reactions
nuclear corrections to parton distributions are at the same level as for electrons, but
have a very different dependence on the Bjorken $x$ variable.
The topic remains controversial, with the hope that future precise Miner$\nu$a results on various
targets will clarify the situation. 

As we already mentioned, the GiBUU code uses PYTHIA for the simulation of DIS processes.
However, the PYTHIA code was designed for elementary reactions. 
In the GiBUU simulation the neutrino interacts with one initial nucleon, bound in the hadronic potential and having nonzero
Fermi momentum. In order to be able to use the PYTHIA event simulator, we have to provide some
quasi-free kinematics as inputs to PYTHIA. Various prescriptions to do this (for details see \cite{Buss:2011mx})
result in a $5-7\%$ difference in the results. 
The corresponding cross sections (denoted as ``F-NO'', ``F-CM'', ``F-THRE'') are shown in Fig~\ref{fig:compare-nuclear}.
Nuclear parton distribution functions
from \cite{Eskola:1998df} are also implemented as one of the options to use 
(to avoid double counting, nuclear potential and Fermi motion in such calculations are switched off). 
The result (``nuclearPDF'') as well as the free cross section for iron composition 
(``26p+30n'') are also shown in Fig~\ref{fig:compare-nuclear}. 
At the moment we consider the various prescriptions mentioned above as intrinsic uncertainty 
of the GiBUU code, reflecting the lack 
of our understanding the nuclear effects. 
No other event generator, as far as we know, accounts for nuclear corrections in high--energy neutrino reactions.  

Fig.~\ref{fig:compare-nuclear} shows, that our calculations are in a good agreement with the recent neutrino data,
which are also consistent with each other.  For antineutrino the agreement is good for $E_{\bar\nu}>25\GeV$.
For lower energies our curve is  above the recent MINOS
data, but below the IHEP-JINR results \cite{Anikeev:1995dj}. The overall agreement of our calculations with the data is
therefore better than the agreement of the data with each other.

\section{Final state interactions and change of the final hadronic spectra.}

\begin{figure}[hbt]
\includegraphics[width=0.7\textwidth]{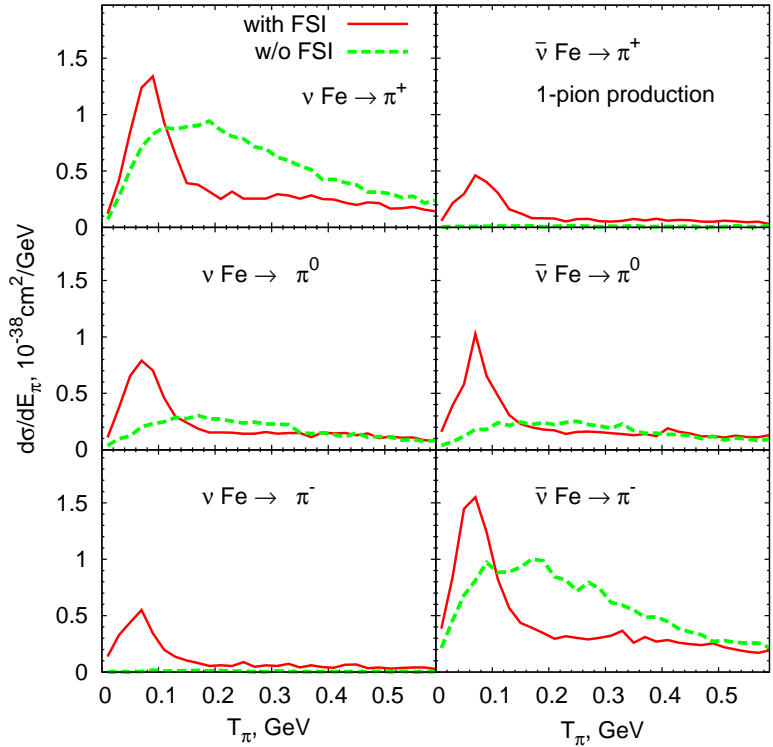}
\caption{(Color online) Pion kinetic energy distributions for neutrino and antineutrino induced reactions for 1-pion production
(one pion of a given charge and no other pions are produced)  ``MULTI'' at least 1 pion of a given charge and any number
of pion of other charges are produced``. Calculations are for the NuMI low-energy mode neutrino/antineutrino fluxes.}
\label{fig:MINOS-ekin-pion-with-FSI-1-pion}
\end{figure}

After being produced in the initial interaction act, outgoing hadrons propagate throughout the nucleus.
In GiBUU this process of final state interactions (FSI) is modeled by solving the semi-classical 
Boltzmann-Uehling-Uhlenbeck equation.
It describes the dynamical evolution of the phase space density for each particle species
under the influence of the mean field potential, introduced in the description of
the initial nucleus state. Equations for various particle species are coupled through this mean field and
also through the collision term. This term explicitly accounts for changes in
the phase space density caused by elastic and inelastic collisions between particles.

FSI decrease the cross sections as well as significantly modify
the shapes of the final particle spectra. Such change was seen, for example,
in photo-pion production~\cite{Krusche:2004uw} and is described by the GiBUU with a good accuracy.
A similar change should be observed in neutrino reactions.

Fig.~\ref{fig:MINOS-ekin-pion-with-FSI-1-pion} shows the $\pi^+$ (upper panels), $\pi^0$ (middle panels)
and $\pi^-$ (lower panels)  spectra for neutrino (left panels) and antineutrino (right panels) NuMI
fluxes for 1-pion events.  The green dashed lines show the kinetic energy ($T_\pi$) distributions without FSI,
i.e. of pions produced in the initial neutrino vertex. The solid red lines show the distributions
after FSI, i.e.  of pions that made it out of the nucleus.
Such spectra can also be calculated for any other predefined final state and should be measurable in Miner$\nu$a experiment.

For dominant channels ($\pi^+$ production for neutrino reactions and $\pi^-$ in antineutrino ones),
the FSI decrease the cross section at $T_\pi > 0.2 \GeV$. This is mainly explained by pion 
absorption through $\pi N \to \Delta$ following by $\Delta N \to NN$. 
Pion elastic scattering in the FSI also decreases the pion energy,
thus depleting the spectra at higher energies and accumulating strength at lower
energies. Thus, an increase of the cross sections is observed at $T_\pi < 0.15 \GeV$;
altogether this leads to a significant change of the shape of the spectra.

Scattering can also lead to pion charge exchange. For neutrino-induced reactions, 
the  $\pi^+ n \to \pi^0 p$ scattering in the FSI is the main source of side--feeding 
for the $\pi^0$ channel, leading to a noticeable increase of the 
$\pi^0$ cross section at low $T_\pi$.  The inverse feeding is suppressed, because less 
$\pi^0$ than $\pi^+$ are produced at the initial vertex. The same mechanism of side feeding
from dominant to sub-dominant channel through $\pi^- p \to \pi^0 n$
is working for antineutrino induced reactions.

For the least dominant channel ($\pi^-$ production for neutrino reactions and $\pi^+$ in antineutrino ones),
the FSI (in particularly side feeding)  represent the main source of the events observed; thus a dramatic 
FSI effect.

\begin{figure}[hbt]
\includegraphics[width=0.8\textwidth]{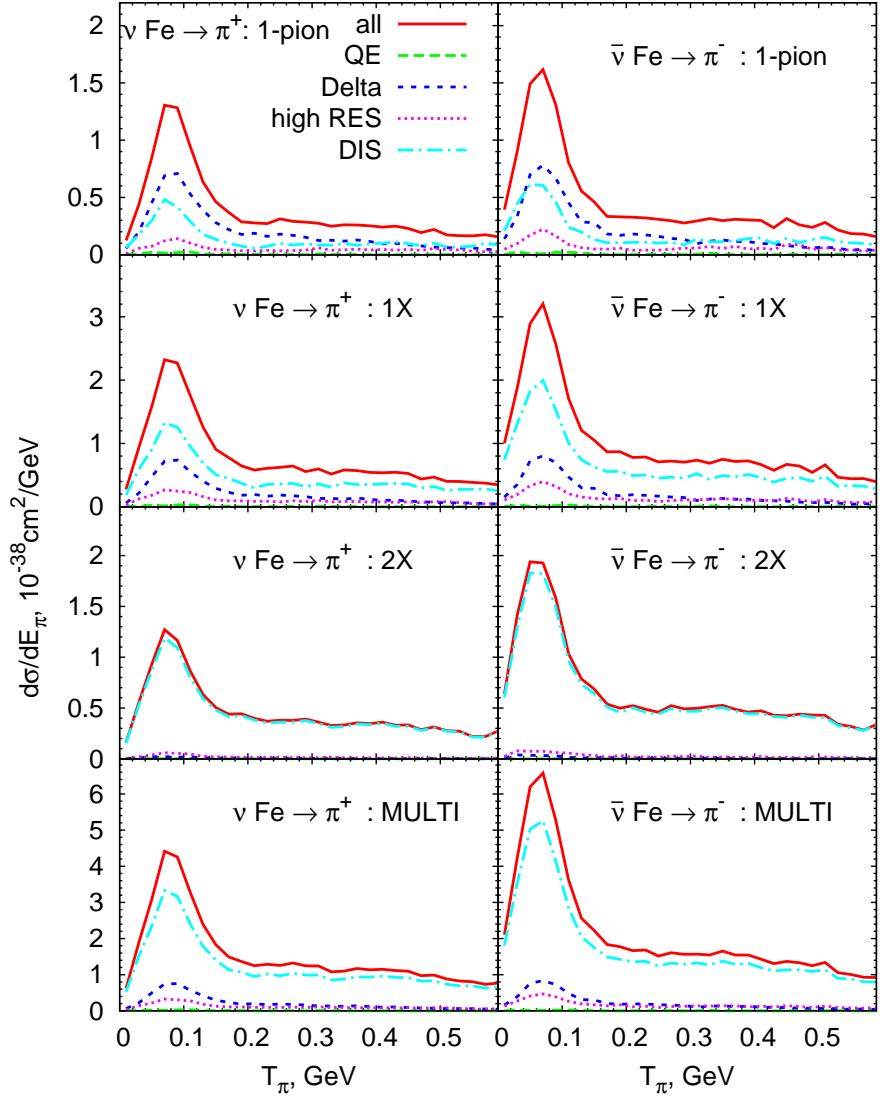}
\caption{(Color online) Pion kinetic energy distributions for $\pi^+$ production in neutrino reaction and $\pi^-$ production in antineutrino reaction
(both are dominant channels), showing various 
contribution to a given final state. 
''1-pion``: one pion of a given charge and no other pions are produced in the final state after FSI; 
''1X``: one pion of a given charge and  any number of other pions;  
''2X``: two pions of a given charge and  any number of other pions;  
``MULTI'': at least 1 pion of a given charge and any number of other pions.
Calculations are for NuMI low-energy mode neutrino/antineutrino fluxes. }
\label{fig:MINOS-ekin-variousOrigins-multiplot-nu-piplus-barnu-piminus}
\end{figure}

Fig.~\ref{fig:MINOS-ekin-variousOrigins-multiplot-nu-piplus-barnu-piminus} shows the origin of the pions 
(that is the initial vertex, at which the pion was produced) in the dominant channels for various final states. 
1-pion production, for example, receives a major contribution comes from Delta resonance production and 
its following decay. For other final states with more pion DIS dominates, but the Delta is still visible.
The contribution from the QE vertex is very small but nonzero, the outgoing pion in this case can only
be produced during the FSI, for example, due to the $NN\to N\Delta$ scattering followed by $\Delta\to N\pi$.

Similar plots can be obtained from the GiBUU simulation for any other outgoing particles (protons, kaons, eta)
for any predefined final state and compared with the coming Miner$\nu$a results.  

\begin{theacknowledgments}
This work is supported by DFG. O.L. is grateful to Ivan Lappo-Danilevski for programming assistance.
\end{theacknowledgments}

\bibliographystyle{aipproc}
\bibliography{nuclear}

\end{document}